# Lunar Palaeoregolith Deposits as Recorders of the Galactic Environment of the Solar System and Implications for Astrobiology

Ian A. Crawford[1,2], Sarah A. Fagents[3], Katherine H. Joy[2,4,5] and M. Elise Rumpf [3]

[1]Department of Earth and Planetary Sciences, Birkbeck College, University of London, Malet Street, London, WC1E 7HX, UK. Email: i.crawford@ucl.ac.uk

[2]Centre for Planetary Sciences at UCL/Birkbeck, University College London, Gower Street, London, WC1E 6BT. UK.

[3]Hawaii Institute of Geophysics and Planetology, University of Hawaii, Honolulu, HI 96822, USA.

[4]The Center for Lunar Science and Exploration, Lunar and Planetary Institute, Houston, TX 77058, USA.

[5]The NASA Lunar Science Institute.

**Abstract.** One of the principal scientific reasons for wanting to resume *in situ* exploration of the lunar surface is to gain access to the record it contains of early Solar System history. Part of this record will pertain to the galactic environment of the Solar System, including variations in the cosmic ray flux, energetic galactic events (e.g, supernovae and/or gamma-ray bursts), and passages of the Solar System through dense interstellar clouds. Much of this record is of astrobiological interest as these processes may have affected the evolution of life on Earth, and perhaps other Solar System bodies. We argue that this galactic record, as for that of more local Solar System processes also of astrobiological interest, will be best preserved in ancient, buried regolith ('palaeoregolith') deposits in the lunar near sub-surface. Locating and sampling such deposits will be an important objective of future lunar exploration activities.

# 1 Introduction

Studies of Apollo samples have revealed that the lunar regolith is efficient at collecting and retaining materials that fall onto it. This includes a record of implanted solar wind material (McKay et al., 1991; Lucey et al., 2006), which may be used to study the composition and evolution of the Sun (e.g Crozaz et al., 1977; Kerridge et al., 1991; Wieler et al., 1996). Recently, samples of the Earth's atmosphere also appear to have been retrieved from lunar regolith samples (Ozima et al., 2005; 2008), opening up the possibility that the lunar geological record may help inform our understanding of the evolution of our own planet. It has further been suggested that samples of Earth's early crust may also be preserved in lunar regolith, in the form of terrestrial meteorites blasted off the Earth in giant impact events (Gutiérrez, 2002; Armstrong et al., 2002; Crawford et al., 2008; see also the paper by J. Armstrong elsewhere in this volume). Indeed, *all* meteorites retrieved from the lunar regolith [of which there have already been several; e.g. McSween (1976), Joliff et al., (1993); Rubin (1997)], would be of compelling interest because of the constraints they would provide on models of the composition and dynamics of near Earth objects throughout Solar System history.

In addition to these local Solar System records, the lunar regolith retains a record of galactic cosmic rays (GCRs) in the form of cosmogenic nuclides and tracks of radiation damage within crystal lattices (e.g. Crozaz et al., 1977; McKay et al., 1991; Goswami, 2001; Lucey et al., 2006), and possibly particles originating in the local interstellar medium (Brilliant et al., 1992; Wimmer-Schweingruber and Bochsler, 2001). All these records would potentially yield valuable information regarding the near-Earth cosmic environment throughout Solar System history, much of it of clear astrobiological interest (e.g. Spudis, 1996; Crawford, 2006; Fagents et al., 2010).

As we have dealt with the importance and preservation of solar wind-implanted volatiles in a previous publication (Fagents et al., 2010), we here concentrate on the record of the Sun's galactic environment that may be preserved in ancient lunar regolith deposits. In Section 2, we first outline the various aspects of this record and its astrobiological significance; in Section 3 we describe how such a record is likely to be preserved on the Moon; and in Section 4 we describe it might be located and recovered; we draw some conclusions in Section 5.

**2 A Lunar Record of the Sun's Galactic Environment**

The Sun has revolved around the Galaxy approximately twenty times since the formation of the Solar System (e.g. Gies and Helsel, 2005), and during this time the Moon's surface will have been exposed to a wide range of galactic environments. At least three consequences of exposure to these different environments are of relevance to astrobiology and may have left a record on the lunar surface: variations in the flux of ionising radiation; occasional accretion of gas and dust during passages through dense interstellar clouds; and variations in the impact cratering rate owing to disturbances of the Oort Cloud. We discuss each of these, and their astrobiological significance, below.

2.1 Variations the flux of galactic ionising radiation

There are several ways in which the flux of ionising radiation (principally GCRs and gamma rays) at the Earth and other terrestrial planets might affect planetary habitability. Possible effects include direct damage to DNA and other biomolecules; planetary cooling due to increased cloud cover owing to tropospheric ionisation and the generation of cloud condensation nuclei; the generation of acid rain resulting from the formation of nitrous oxides; and increased surface UV flux as a result of cosmic ray-induced ozone destruction (for recent reviews covering these various processes see: Shaviv, 2006; Scherer, et al., 2006; Kirby and Carslaw, 2006; Medvedev and Melott, 2007; Ferrari and Szuszkiewicz, 2009; and Dartnell, 2010).

As summarised by Shaviv (2006) and Scherer et al. (2006), there are a variety of galactic processes that affect the GCR flux in the inner Solar System, operating on a range of timescales. On the longest timescales (> 1 Gyr), the average GCR flux probably reflects the galactic star formation rate. This would provide very useful constraints on models of galactic evolution, but is perhaps only of tangential interest to astrobiology (although, of course, a better understanding of star formation rates, and associated galactic chemical evolution, is relevant to understanding the history of galactic habitability on the largest spatial and temporal scales). Of more direct astrobiological interest would be to search for evidence of very energetic events (e.g. gamma-ray bursts; GRBs), which may have been more frequent earlier in galactic history and which potentially could have had a drastic effect on habitable planets throughout the Galaxy (e.g. Thorsett 1995; Annis, 1999; Scalo and Wheeler, 2002; Thomas, 2009). Although life on Earth appears never to have been exterminated by such events, it would nevertheless be of interest to know whether any occurred which might be correlated with evolutionary events in the fossil record (and, again, quantifying the frequency of GRBs would improve our understanding of galactic habitability as a whole). While the consequences of a galactic GRB on the terrestrial environment may have been severe, the Earth's surviving geological record from the time period in question is so poorly preserved that identifying an unambiguous record of such events on this planet appears unlikely; the lunar surface, and in particular sufficiently ancient and well-preserved regolith deposits, would seem to offer much stronger prospects.

On shorter timescales (of the order of a few 100 Myr) the GCR flux is likely to be moderated by the enhanced supernova rate associated with the Sun passing through galactic spiral arms (e.g. Leitch and Vasisht 1998; Shaviv, 2003; Gies and Helsel, 2005). This raises the intriguing possibility that the record of GCR fluxes retained in the lunar regolith might be able to help constrain models of the Galaxy's spiral structure, which is of inherent astronomical interest. Of particular interest from an astrobiological point of view would be occurrences of nearby ($\leq 8$ parsecs) supernova explosions, which would be expected to result in serious biological consequences (e.g. Gehrels et al., 2003). Fortunately, such 'point blank' supernovae are unlikely to be encountered in every spiral arm crossing: Gehrels et al. (2003) estimate that one may be expected only every 1.5 Gyr or so, which is a time interval much better matched to the lunar geological record than the terrestrial one. Identifying spiral arm passages, and nearby supernovae explosions, from the terrestrial biological and climate records is controversial because of the ambiguity and sparseness of the data (see e.g. Rahmstorf et al, 2004), and it is precisely in this respect that access to the longer duration, and almost certainly better preserved, lunar GCR record would help.

On still shorter timescales (tens of Myr), additional variations in the GCR flux may be expected owing to the oscillation of the solar orbit about the plane of the Galaxy. Currently, the amplitude of this oscillation is about 70 parsecs, with a period of about 64 Myr (Gies and Helsel, 2005). At first sight, one might expect a lower cosmic ray flux when the Sun is located above or below the galactic plane, both because of the increased distance from probable supernova sites in the plane and because the lower density of the surrounding interstellar medium will cause the heliosphere to expand and result in greater shielding of the inner Solar System (e.g. Shaviv, 2003). However, Medvedev and Melott (2007) have recently argued that, owing to the Galaxy's motion through the local inter-galactic medium, the Solar System is in fact exposed to an enhanced flux of *extragalactic* cosmic rays when at the northern extremity of its oscillations about the galactic plane. Medvedev and Melott (2007) correlate this effect with an apparent 62±3 Myr periodicity in fossil diversity (Rohde and Muller, 2005). They argue that biodiversity appears lower when the extragalactic cosmic ray flux is highest (i.e. when the Sun is well above the galactic plane), and that this correlation implies a causal connection. As for the attempts to correlate terrestrial events with supernova explosions and spiral arm passages described above, confirming a direct link between variations in the cosmic ray flux and evolutionary events on Earth requires access to a better record of the former, and suitably preserved lunar regolith deposits may help in this respect.

2.2 Accretion of gas and dust from interstellar clouds

Currently, the Sun lies within a very low density region of the interstellar medium (with a hydrogen density, $n_H$ ~ 0.2 cm$^{-3}$; e.g. Frisch et al., 2009). However, during its orbit about the Galaxy, and especially during spiral arm passages, it will have encountered much denser interstellar environments (Talbot and Newman, 1977; Shaviv, 2003; Gies and Helsel, 2005). In addition to the enhanced supernova rate discussed above, this raises the possibility that material from dense interstellar clouds might be directly accreted onto the atmospheres and surfaces of terrestrial planets, with possible biological consequences (e.g. Yabushita and Allen, 1989; Pavlov et al., 2005; Yeghikyan and Fahr, 2006).

The principal effect of the Solar System entering a dense interstellar cloud is for the heliosphere to shrink as the pressure of the solar wind balances a higher external pressure (Yeghikyan and Fahr, 2006; Smith and Scalo, 2009). In a review of this process, Yeghikyan and Fahr (2006) found that for interstellar cloud densities of $n_H \geq 1000$ cm$^{-3}$ the radius of the heliosphere shrinks to less than 1 AU, leaving the Earth (and Moon) directly exposed to interstellar material; a more recent study by Smith and Scalo (2009) found a similar, but somewhat smaller, critical density of $n_H \geq 600$ cm$^{-3}$. There is little doubt that the Solar System will have encountered such interstellar clouds during its history, but estimates of the associated timescales differ. Talbot and Newman (1977) estimated that the Sun would encounter such clouds approximately every 300 Myr on average, whereas Yeghikyan and Fahr (2006) obtained a comparable, but somewhat longer, interval of 460 Myr. On the other hand, Scoville and Sanders (1986) estimated much longer intervals (~ 1 Gyr) between such encounters. This range is bracketed by the recent analysis of Smith and Scalo (2009), who argued that based on current knowledge it is not possible to determine the frequency of such encounters to better than an order of magnitude, but that they probably occur somewhere between 100 Myr and 1 Gyr.

Pavlov et al. (2005) made a number of suggestions for trying to identify such events in Earth's geological record (for example by looking for non-terrestrial $^{235}$U/$^{238}$U ratios, where fresh interstellar material will have a higher abundance of $^{235}$U owing to its continual production by supernovae); Shaviv (2006) advanced a similar argument based on a search for $^{244}$Pu. Given the likely distinctive chemical and isotopic composition of interstellar dust, there would presumably be other geochemical signatures for which one might search. The calculations of Pavlov et al. (2005) imply that a 'typical' interstellar cloud crossing ($n_H \sim$ 1000 cm$^{-3}$ for 200,000 years) would deposit ~ 1 kg m$^{-2}$ of interstellar dust on exposed planetary surfaces. As for the other evidence of spiral arm crossings discussed above, the stable surface of the Moon would seem better suited to retaining such a record than the terrestrial one, especially if the timescales involved turn out to be at the longer end of the range identified by Smith and Scalo (2009).

In addition to the collection of interstellar dust, there is also the possibility of identifying the gaseous component of the local interstellar medium in lunar soils, perhaps as interstellar pickup ions (Wimmer-Schweingruber and Bochsler, 2001). Note that, owing to the mechanism of implantation, this is a record which will only be preserved in the surficial regolith of an airless body such as the Moon. Wimmer-Schweingruber and Bochsler (2001) have suggested that interstellar pick-up ions may in fact already have been identified in lunar soils, although more recently it has been argued that the isotopic profile originally identified as such may actually be an artifact due the preferential sputtering of the outer layers of the grains studied (e.g. Wieler et al., 2006). Nevertheless, in principle interstellar pickup ions may still be present within lunar regolith grains, albeit harder to identify than originally thought (Wimmer-Schweingruber, personal communication). A careful study of these would reveal much about local galactic environment, including periods of dense cloud crossings and heliosphere collapse.

From an astrobiological perspective, direct exposure of the Earth's atmosphere to an interstellar cloud with a density of $n_H \sim 1000$ cm$^{-3}$, which would probably be sustained for several hundred thousand years, would have a number of consequences for habitability. In particular, the influx of hydrogen would alter atmospheric chemistry in such a way as to deplete mesospheric ozone, and the resulting change in temperature profile may lead to the formation of world-wide mesospheric clouds leading to a global decrease in temperature (Yeghikyan and Fahr, 2006). The direct accretion of interstellar dust could further increase atmospheric opacity and contribute to additional cooling (Pavlov et al., 2005). In addition, collapse of the heliosphere to within the Earth's orbit would lead to an increased GCR flux, with the additional consequences discussed in Section 2.1 above.

Although the major astrobiological interest in the history of encounters between the Solar System and dense interstellar clouds are the consequences for planetary habitability, there are other possibilities. For example, Napier (2007) has drawn attention to the possible role of dense interstellar clouds in facilitating the transfer of life between planetary systems (i.e. interstellar panspermia). Here, the suggestion is that life-bearing meteorites ejected from planetary systems (e.g. Melosh, 2003) may become concentrated within molecular clouds, from which they may later be accreted by other planetary systems passing through the cloud. Historical records of the Solar System's own passages through such clouds, and the composition of any interstellar dust collected, could provide observational constraints on theories of this kind.

2.3 Variations of the impact cratering rate

In addition to experiencing variations in the intensity of ionizing radiation and local interstellar densities, the Sun also experiences a changing *gravitational* environment as it orbits the Galaxy. In particular, the vertical oscillation above and below the plane, and passages through spiral arms (with their attendant close passes to massive molecular clouds) causes variations in the gravitational potential which may perturb the orbits of comets in the Oort cloud and periodically increase the cratering rate in the inner Solar System (e.g. Torbett, 1986; Clube and Napier, 1986; Matese et al., 1995; Stothers, 1998; Leitch and Vasisht, 1998). Such variations in the impact rate are of interest to astrobiology because of their potential consequences for the habitability of planetary surfaces. However, attempts to correlate the terrestrial impact rate with variations in biodiversity are hampered by the relative sparseness of the terrestrial impact record. The Moon, on the other hand, retains an essentially complete record for impacts of all sizes over the whole of Solar System history (e.g. Stöffler et al., 2006), making this the most logical place to confirm or refute such correlations.

The lunar record will be particularly valuable in correlating variations in impact rate with variations in cosmic ray intensity. Given sufficiently wide-ranging sampling, the density of craters on well-dated surfaces will yield the former, while the cosmic ray exposures of regoliths developed on, or buried by, such surfaces will yield the latter. Some putative galactic influences on the Solar System (e.g. spiral arm crossings; Leitch and Vasisht, 1998) would predict a positive correlation between these variables, while others (e.g. exposure to extra-galactic cosmic rays during the Sun's excursions above the plane; Medvedev and Melott, 2007) would predict an anti-correlation (with a different period). Access to the long-lived, and well-preserved, lunar records of *both* impact history and cosmic-ray flux may be the only way to fully disentangle these competing processes.

## 3. Palaeoregolith Deposits: Preserving a Record

We have argued here that the lunar surface, and especially the lunar regolith, will contain a uniquely valuable record of the galactic environment of the Sun, and that this record will be of importance to the science of astrobiology (as well as galactic astronomy and heliospheric physics). However, the present surficial regolith has been subject to comminution and overturning ("gardening") by meteorite impacts for the last three to four Gyr, and the record it contains is therefore an average over most of Solar System history. This will complicate the retrieval and interpretation of any record it may contain (as also pointed out by Wimmer-Schweingruber and Bochsler, 2001). From the point of view of obtaining ancient records of solar system and galactic history it will be most desirable to find ancient regoliths (*palaeoregoliths*) that have been undisturbed since their formation.

A regolith forms when a fresh lunar surface is exposed to the flux of micrometeorites. Most exposed mare surfaces date from between ~ 3.8 and 3.1 Gyr ago, with small-scale, geographically restricted volcanism continuing to perhaps as recently as 1 Gyr ago (Hiesinger et al., 2003). For example, the study by Hiesinger et al. (2003; cf. their Fig. 10) reveals a patchwork of discrete lava flow units in Oceanus Procellarum with individual ages ranging from about 3.5 to 1.2 Gyr ago. As younger lava flows are superimposed on older ones, we may expect to find layers of palaeoregoliths sandwiched between lava flows dating from within this age range (see Fig. 1). As discussed by Crawford et al. (2007) and Fagents et al. (2010), the archival value of such palaeoregoliths will be enhanced by the fact that the under- and overlying basalt layers will lend themselves to radiometric dating, thereby defining the age of the geological record sandwiched between them.

A worthwhile geochemical record will only be preserved within a palaeoregolith layer if it survives the thermal consequences of burial by the initially molten overlying lava flow. In previous work (Fagents et al., 2010) we have shown that, for lava flows ranging from 1 to 10 m thickness, implanted solar wind particles should be substantially preserved in palaeoregoliths at depths of greater than 40 to 400 mm beneath an overlying lava flow, respectively. The rate of regolith accumulation varies as a function of exposed surface age (as the growing regolith itself impedes further regolith development), and with absolute age (owing to higher impact rates in the past). Horz et al. (1991) give regolith growth rates in the range 1 to 5 mm $Myr^{-1}$, where the lower value is the current rate and the higher value refers to fresh basaltic surfaces about 3.8 Gyr ago. A value of 2 mm $Myr^{-1}$ would seem to be conservative for fresh basalt surfaces over most of lunar history, which implies that individual lava flows would have to remain exposed for between 20 and 200 Ma to accumulate regoliths in the thickness range required to substantially shield implanted volatiles from an overlying lava flow. The ages of individual basalt flows mapped by Hiesinger et al. (2003) indicate that this is likely to have been a common occurrence.

For the purposes of this paper, we are primarily concerned with the preservation of cosmic ray records (i.e. cosmogenic nuclei and radiation tracks), interstellar pickup ions, and geochemical evidence for interstellar dust particles, rather than the solar wind particles considered by Fagents et al. (2010). However, it seems likely that in general these will be more robust against thermal disruption than the solar wind particles. For example, the interaction of cosmic rays with the regolith, and therefore the production of cosmogenic nuclei, occurs at much deeper levels [i.e. of the order of a meter; Vaniman et al. (1991), Goswami (2001)] than the solar wind implantation (typically less than a micron). Moreover, with the exception of $^3$He, most cosmogenic nuclei that might be used to measure the cosmic ray flux [e.g. $^{21}$Ne, $^{38}$Ar, $^{83}$Kr and $^{126}$Xe; Goswami (2001), Eugster (2003)] are relatively heavy and would therefore only be expected to outgas at high temperatures (see Fig. 2 of Fagents et al., 2010). We note that high energy GCRs are likely to pass through regolith layers less than about a meter thick and penetrate the underlying bedrock, but this will not affect the survival of a GCR-induced record provided that the overlying regolith is sufficiently thick to attenuate the thermal wave when the overlying lava flow is emplaced. Cosmic ray tracks would also be expected to occur at depths greater than the likely penetration of the thermal wave (although their sensitivity to thermal annealing remains to be investigated). Any interstellar pickup ions (such as those tentatively identified by Wimmer-Schweingruber and Bochsler, 2001), would be expected to be found at depths comparable to the solar energetic particles [i.e. on the order of mm; Vaniman et al. (1991), Goswami (2001)], and are therefore likely to be less susceptible to thermal desorption than the solar wind implanted ions. Finally, the thermal effects of an overlying lava flow on the geochemical signature of accreted interstellar dust is not known, but as such identification is unlikely to depend on the detection of highly volatile substances there is no obvious reason to think that this would be a serious obstacle.

Detailed studies of all these effects will be the topic of future work, but it seems most likely that the galactic records to be found within lunar palaeoregoliths will be no more, and probably significantly less, susceptible to the thermal consequences of burial by lava flows than the solar wind implanted particles discussed by Fagents et al. (2010).

## 4. Locating and Accessing Palaeoregolith Deposits

By definition, palaeoregolith deposits will be located below the surface, although surface mapping of lava flows of different ages (e.g. Hiesinger et al., 2003; Bugiolacchi et al., 2006; Hackwill, 2010) will help identify areas where such deposits are likely to be found (see the maps presented in these papers as a guide to likely locations). High-resolution imaging, such as that being obtained with the Lunar Reconnaissance Orbiter Camera (LROC; Robinson et al., 2010) will be especially valuable in this respect (e.g. Crawford et al., 2009). Studies of areas where small impact craters have excavated through overlying lava flows to reveal sub-surface boundaries (e.g. Weider et al., 2010), and/or orbital ground penetrating radar measurements (e.g. Ono et al., 2009), will help confirm the likely occurrence of palaeoregolith layers and provide some information on their depth and thickness.

Ultimately, we will need to identify specific localities where buried palaeoregoliths may be accessible to future robotic or human exploration. Note that, while LROC images may identify candidate palaeoregolith outcrops in the walls of rilles or large craters, these are likely to be relatively inaccessible. In addition, such outcrops are unlikely to yield undisturbed palaeoregolith deposits; impact cratering, in particular, could seriously disrupt the record we are seeking. Ideally, therefore, it would be desirable to identify localities where modest drilling (e.g. to depths of a few tens of meters) would permit access to undisturbed palaeoregolith deposits. Although initial exploratory sampling might be achieved robotically, we argue that gaining full access to the lunar palaeoregolith record (as for other areas of lunar exploration) would be greatly facilitated by a renewed human presence on the lunar surface.

The requirements of an exploration architecture able to locate and access lunar palaeoregolith deposits has been outlined by Crawford et al. (2007), and, briefly, would consist of the following elements:

- The ability to conduct 'sortie-class' expeditions to a range of localities on the lunar surface;
- Provision for surface mobility – in the specific case of the Procellarum basalt flows mapped by Hiesinger et al. (2003), a range of order 250 km would permit access to a number of different units with a wide range of ages. In the context of a human mission this implies provision of a pressurized rover;
- Provision of the means to detect sub-surface palaeoregolith deposits. A rover-borne ground penetrating radar system (e.g. Heggy et al., 2009) would be a suitable means of achieving this;
- Provision of a drilling capability, perhaps to ~ 100 m depths (for a review of suitable planetary drilling technology see Zacny et al., 2006). This also implies provision for storage and transport of the resulting drill cores; and

- Adequate provision for sample collection and return capacity (roughly estimated at several 100 kg per sortie).

## 5. Conclusions

Lunar palaeoregolith deposits potentially contain important records of the evolution of the Sun, and its galactic environment, throughout most of Solar System history (approximately 20 revolutions of the Galaxy). In particular, palaeoregoliths likely contain records of variations in the cosmic ray flux, energetic galactic events (e.g, supernovae and gamma-ray bursts), and passages of the Solar System through dense interstellar clouds. Much of this record is of astrobiological interest as these processes may have affected the evolution of life on Earth, and perhaps other Solar System bodies. Locating and sampling such palaeoregolith deposits will be an important objective of future lunar exploration activities. As this will require extensive field excursions, probably combined with sub-surface drilling, we argue that it would be greatly facilitated by a renewed human presence on the Moon

## Acknowledgements

We would like to thank Robert Wimmer-Schweingruber for helpful correspondence regarding the possibility of identifying interstellar pickup ions in lunar regolith grains, and two anonymous referees for helpful comments. SAF and MER acknowledge support from the NASA Lunar Advanced Science and Exploration Research (LASER) program through grant NNX08AY75G and NSF-GRFP.

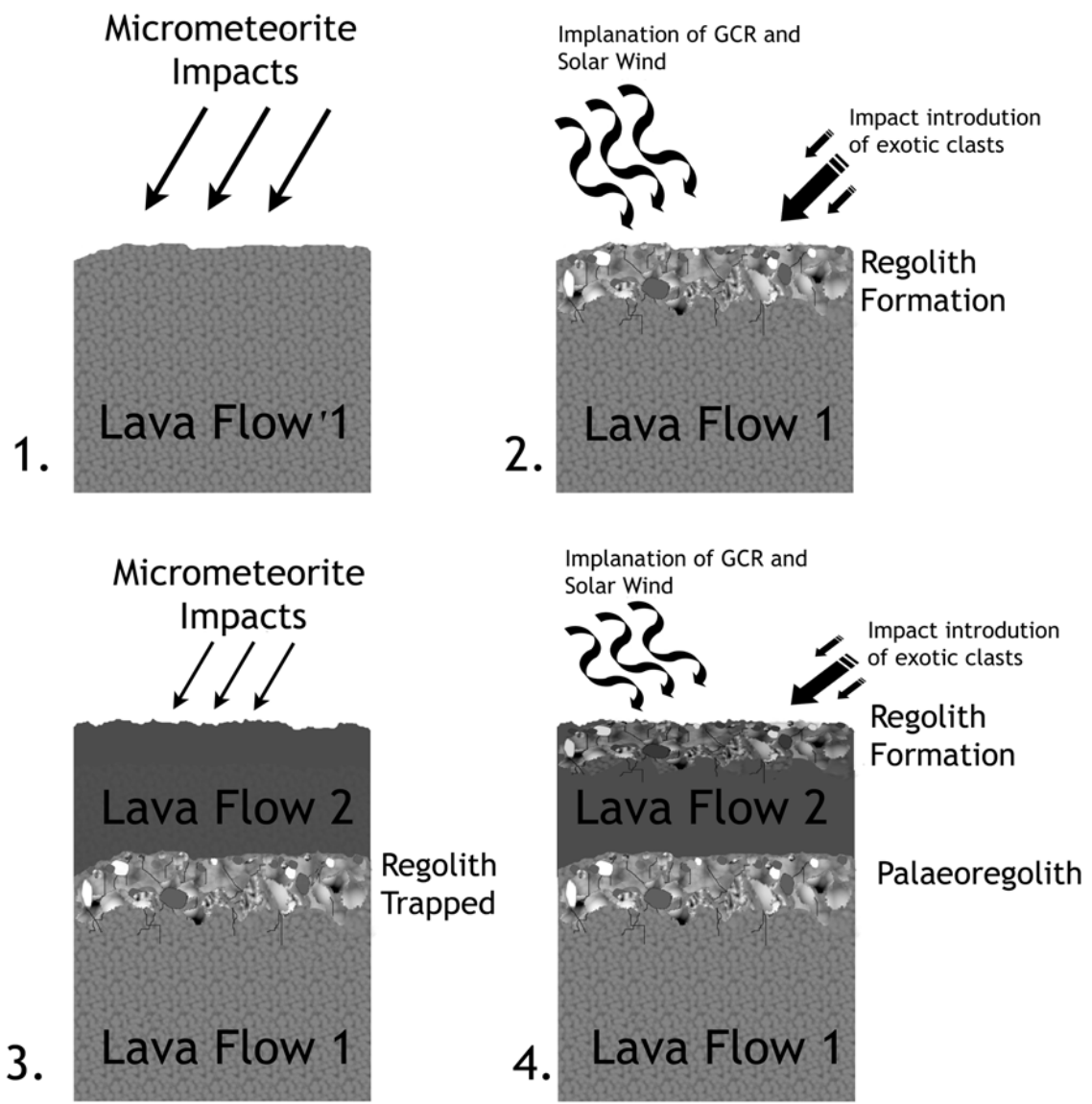

**Figure 1.** Schematic representation of the formation of a palaeoregolith layer: (1) a new lava flow is emplaced and meteorite impacts immediately begin to develop a surficial regolith; (2) solar wind particles, galactic cosmic-rays, and "exotic" material derived from elsewhere on the Moon (and perhaps elsewhere, such as the early Earth) are implanted; (3) the regolith layer, with its embedded historical record, is buried by a more recent lava flow, forming a palaeoregolith; (4) the process begins again on the upper surface. (Image reproduced from Crawford et al. (2007) with permission; © Royal Astronomical Society).